\documentclass[9pt,conference]{IEEEtran}
\usepackage[preprint]{waspaa25}

\usepackage{bm} % for bold math symbols (incl. Greek letters) with \bm{}

\title{Learning to Upsample and Upmix Audio in the Latent Domain}

\name{Dimitrios Bralios$^{1,\star}$\thanks{* Work done as an intern at Adobe Research.\newline{Demos: \texttt{https://re-encoders.github.io/latent-bwe-m2s/}}},
      Paris Smaragdis$^{1}$,
      Jonah Casebeer$^{2}$}
\address{$^{1}$University of Illinois Urbana-Champaign, Urbana, IL, USA \\
$^{2}$Adobe Research, San Francisco, CA, USA
}

\begin{document}

\maketitle

\begin{abstract}
Neural audio autoencoders create compact latent representations that preserve perceptually important information, serving as the foundation for both modern audio compression systems and generation approaches like next-token prediction and latent diffusion. Despite their prevalence, most audio processing operations, such as spatial and spectral up-sampling, still inefficiently operate on raw waveforms or spectral representations rather than directly on these compressed representations. We propose a framework that performs audio processing operations entirely within an autoencoder's latent space, eliminating the need to decode to raw audio formats. Our approach dramatically simplifies training by operating solely in the latent domain, with a latent L1 reconstruction term, augmented by a single latent adversarial discriminator. This contrasts sharply with raw-audio methods that typically require complex combinations of multi-scale losses and discriminators. Through experiments in bandwidth extension and mono-to-stereo up-mixing, we demonstrate computational efficiency gains of up to 100× while maintaining quality comparable to post-processing on raw audio. This work establishes a more efficient paradigm for audio processing pipelines that already incorporate autoencoders, enabling significantly faster and more resource-efficient workflows across various audio tasks.
\end{abstract}

% \begin{IEEEkeywords}
% latent audio processing, bandwidth extension, mono-to-stereo, efficient processing.
% \end{IEEEkeywords}

\section{Introduction}

% Neural audio autoencoders are key to many common tasks now
Neural audio autoencoders have become central to modern audio technologies, adept at creating compact latent representations that preserve perceptually important information~\cite{zeghidour2021soundstream, defossez2023high, kumar2024high, caillon2021rave}. These representations form the bedrock of many audio compression systems and enable powerful generative models, including text-to-audio synthesis via latent diffusion~\cite{liu2023audioldm, huang2023make, evans2024stable, evans2024fast, zhu2025review, liu2024audioldm}, next token prediction~\cite{kreukaudiogen, copet2023simple, borsos2023audiolm, agostinelli2023musiclm, yang2024uniaudio} and  parallel decoding~\cite{borsos2023soundstorm, ziv2024masked, flores_garcia2023vampnet}. Despite the prevalence of autoencoders, many fundamental audio processing operations, such as bandwidth extension and mono-to-stereo upmixing, are still predominantly performed on raw waveform or spectral representations~\cite{mandel2023aero, zhu2024musichifi, moliner2022behm, moliner2024blind, li2025apollo}.

% For some reason audio-audio tasks have missed out
This reliance on raw audio processing introduces significant inefficiencies, particularly within pipelines that already utilize an autoencoder, whether for transmission or generation. First, it necessitates decoding the latent representation back to its high-dimensional raw format~(waveform or spectral) simply to apply these operations, incurring overhead especially if the result must then be re-encoded. Second, the processing models operate on raw audio, likely learning to construct and manipulate low-level acoustic features already captured and represented by the autoencoder.

% We address this gap
To address these limitations, we propose the \emph{Re-Encoder} framework. This setup performs audio processing operations entirely within a pre-trained autoencoder's latent space, during both training and inference. A key advantage of our approach is its autoencoder-agnostic nature. It is designed to operate on the latent representations produced by any pre-trained audio autoencoder, directly leveraging existing research and development in this area. Essentially, the Re-Encoder re-uses the original autoencoder's encoder. This approach fundamentally eliminates the need to decode to raw audio for intermediate processing, directly leveraging the compact and informative nature of latent representations. Operating solely in the latent domain dramatically simplifies the training process. Our method employs a straightforward latent $L_1$ reconstruction objective, optionally augmented by a single latent adversarial discriminator. This contrasts sharply with conventional raw-audio methods, which typically demand complex combinations of multi-scale spectral, mel, or waveform losses, along with multiple discriminators operating on raw audio, significantly increasing training complexity, cost, and potentially hindering stability~\cite{schwar2023multi}.

\begin{figure}[t]
    \centering
        \centering
        \includegraphics[width=.95 \columnwidth, trim=0pt 0pt 15pt 0pt, clip]
{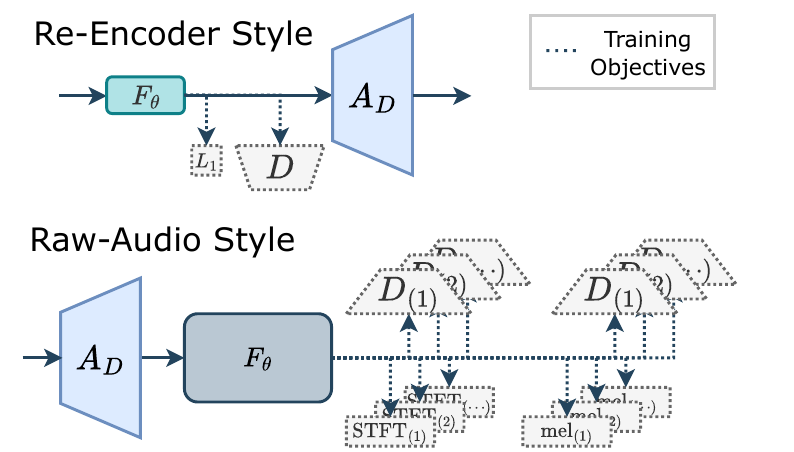}
        % {fig/Fig1-full-7.pdf}
        % {fig/Fig1_re_enc.drawio.pdf}
        % {fig/Fig1_2.drawio.pdf} %{fig/Fig1_bwe.drawio-2.pdf}
        \caption{\textbf{Bottom:} Conventional post-generation/transmission workflow—decode to raw audio, then process it. Training involves multiple reconstruction and adversarial objectives.
\textbf{Top:} Our proposed method—perform all processing directly on the compact latent representation before the decoding step, avoiding costly operations on raw audio. Training performed solely in the latent domain.}
    \label{fig:fig1}
    %\vspace{-0.5em}
\end{figure}

% Some others have addressed parts of this gap
While prior work has explored audio processing within latent spaces, most methods employ computationally intensive large models. For example, bandwidth extension has primarily used diffusion-based approaches on continuous latents~\cite{liu2024audiosr} or discrete tokens~\cite{fang2024vector}, and token-based transformers have been applied to source separation~\cite{tokensplit, yang24h_interspeech}. While more efficient separation methods exist for unquantized neural audio codec latents~\cite{yip2024towards, li2025speech, yip2024speech, zhu2025review}, their high-dimensional representations hinder efficiency. Similarly, recent work repurposes neural audio codecs as mel-vocoders~\cite{lanzendorfer2025high} or speech enhancers~\cite{casebeer2021enhancing}, but this requires fine-tuning the base autoencoder with expensive and complex time-domain objectives. In contrast, our approach tackles standard audio tasks using lightweight modules that operate directly on highly compressed latents produced by off-the-shelf, pre-trained audio autoencoders (achieving 16$\times$ or greater compression).

% Facts about how we addresed it
In this work, we demonstrate the effectiveness of our latent-domain processing framework through experiments on bandwidth extension and mono-to-stereo up-mixing. We demonstrate that operating entirely within the latent space allows for training models in less than two days on a single GPU. At inference, this approach achieves 100$\times$ efficiency gains while maintaining performance comparable to raw-audio post-processing techniques. This research establishes a more efficient and resource-conscious paradigm for audio processing, particularly beneficial for pipelines already incorporating neural audio autoencoders, thereby 
enabling faster, more integrated workflows across a variety of audio tasks. 
Crucially, our autoencoder-agnostic framework is positioned to benefit from future autoencoder advancements. 

%Crucially, the autoencoder-agnostic nature of our framework ensures that it is positioned to benefit from future advancements in audio autoencoder technology. 

\section{Proposed Method}
\label{sec:method}

\subsection{Latent Space Audio Processing}
% We will operate on pretrained autoencoders
Drawing inspiration from the compact and information-rich latent spaces learned by modern neural audio autoencoders, our framework operates on these latents. We ``Re-Encoder" a pre-trained and frozen audio autoencoder, comprising an encoder $A_E : \mathbb{R}^{L} \to \mathbb{R}^{C \times T}$ and a decoder $A_D: \mathbb{R}^{C \times T} \to \mathbb{R}^{L}$. The encoder maps a high-dimensional input waveform $\mathbf{x} \in \mathbb{R}^{L}$ to a lower-dimensional sequence of latent vectors $\mathbf{z} =A_E(\mathbf{x}) \in \mathbb{R}^{C \times T}$, where the latent dimensionality ${C \times T \ll L}$. The decoder $A_D$ is designed to reconstruct the original waveform, $\hat{\mathbf{x}} = A_D(\mathbf{z}) \in \mathbb{R}^{L}$,  from this latent representation. 

% We will use latent targets
For audio-to-audio tasks, our objective is to predict the target latent representation $\mathbf{z}_\mathrm{tgt}=A_E(\mathbf{x}_{\mathrm{tgt}})$ given an input latent representation $\mathbf{z}_\mathrm{in}=A_E(\mathbf{x}_{\mathrm{in}})$ and optional conditioning information, $\mathbf{c} \in \mathbb{R}^H$. We train a task specific predictive model ${F}_\theta$, which acts as a mapping ${F}_\theta: \mathbb{R}^{C \times T} \times \mathbb{R}^H \to \mathbb{R}^{C \times T}$, estimating the target latent as $\hat{\mathbf{z}}_\mathrm{tgt} = {F}_\theta(\mathbf{z}_\mathrm{in}, \mathbf{c})$. By operating on the compressed latent representation, the model ${F}_\theta$ can be significantly more lightweight in terms of parameter count and computational requirements compared to models operating directly on the high-dimensional waveform $\mathbf{x}$. Consequently, inference with ${F}_\theta $ is highly efficient.

% We don't backprop through the decoder OR the encoder
Training the predictive model ${F}_\theta$ directly in the waveform space presents a significant challenge, primarily requiring costly forward and backward passes through the decoder for loss computation. Furthermore, optimizing waveform-based objectives often necessitates intricate combinations of multi-scale losses and multiple adversarial discriminators, demanding extensive hyperparameter tuning~\cite{schwar2023multi, kumar2019melgan, kumar2024high, zhu2024musichifi}. To circumvent these inefficiencies and complexities, we propose training  ${F}_\theta$ using loss functions defined exclusively in the latent space. This approach eliminates the need for decoder passes during training, requiring only encoder passes to obtain the latent representations of the input and target waveforms. The primary objective optimized during training is the $L_1$-distance between the predicted latent and the target latent,
\begin{equation}
\mathcal{L}_{\mathrm{rec}} = \mathbb{E}_{\mathbf{x}_\mathrm{in}, \mathbf{x}_\mathrm{tgt}, \mathbf{c}} \left[ \left\| A_E(\mathbf{x}_\mathrm{tgt}) - {F}_\theta (A_E(\mathbf{x}_\mathrm{in}), \mathbf{c}) \right\|_1 \right],
\end{equation}
where $\| \cdot \|_1$ denotes the $L_1$-distance, $\mathbf{x}_\mathrm{in}$ are the $\mathbf{x}_\mathrm{tgt}$ are the input and target waveforms, and $\mathbf{c}$ is the condition. Minimizing this $L_1$ distance aligns predicted and target vectors by leveraging the latent space's property: proximity corresponds to waveform perceptual similarity. Empirically, we find this simplifies optimization compared to directly minimizing waveform perceptual errors.

% Here are the two extensions we experiment with
To demonstrate the flexibility and generality of the Re-Encoder paradigm, we explore two distinct extensions. First, we introduce a discriminator operating directly on the latent representations to align the distribution of predicted latents with that of target latents, which we evaluate using the task of bandwidth extension. Then, we explore variational conditioning~\cite{yang2022upmixing} in the latent space, enabling the modeling of a distribution over the target latent space, which we evaluate on the task of spatially conditioned mono-to-stereo conversion.

\subsection{Latent Space Discriminator for Bandwidth Extension}
Building upon the advantages of operating within the compact and perceptually meaningful latent space, we propose incorporating a discriminator directly into this latent domain. Traditional generative models operating on high-dimensional data like audio often employ adversarial training with discriminators that evaluate outputs in the original data space (e.g., waveform discriminators). However, training such discriminators can be computationally expensive, require complex multi-scale architectures, and would necessitate backpropagation through the decoder during generator training. In contrast, a latent discriminator $D: \mathbb{R}^{1 \times C \times T} \to \mathbb{R}^{1 \times C^\prime \times T^\prime}$, could be designed to operate solely on the latent representation, attempting to distinguishing between latents which came from the ground truth or predictive model distributions. This encourages ${F}_\theta$ to resolve uncertainty in its estimates in a manner consistent with the target distribution.

% Paragraph about BWE
We evaluate this latent discriminator approach on the task of Bandwidth Extension~(BWE). For BWE, our model receives as input audio with a low sampling rate and aims to produce a full-band estimate. Specifically, we first upsample the input low-bandwidth waveform $\mathbf{x}_\mathrm{in}$ using sinc interpolation, and then extract  $\mathbf{z}_\mathrm{in} = A_E(\mathbf{x}_\mathrm{in})$, which are fed as input to ${F}_\theta$. The target latent vectors $\mathbf{z}_\mathrm{tgt} = A_E(\mathbf{x}_\mathrm{tgt})$ are obtained from the corresponding full-band target waveform $\mathbf{x}_\mathrm{tgt}$. Following recent work in audio generation employing adversarial training~\cite{ kumar2019melgan, kumar2024high, zhu2024musichifi}, we utilize a combination of a least-squares GAN loss~\cite{mao2017least} and a feature matching loss component~\cite{kumar2019melgan}.

\begin{align}
\mathcal{L}^{{F}}_{\mathrm{adv}} &= \mathbb{E}_{\mathbf{z}_\mathrm{in}} \left[ \left(1 - D({F}_\theta(\mathbf{z}_\mathrm{in})\right)^2
\right],\\
\mathcal{L}^{{F}}_\mathrm{fm} &=  \mathbb{E}_{\mathbf{z}_\mathrm{in}, \mathbf{z}_\mathrm{tgt}} \left[\sum_{i=1}^N \frac{\| D^i (\mathbf{z}_\mathrm{tgt}) - D^i({F}_\theta(\mathbf{z}_\mathrm{in}))\|_1}
{\|
D^i({F}_\theta(\mathbf{z}_\mathrm{in}))
\|_1}
\right],
\end{align}
where $D^i$ denotes the discriminator feature map at the $i$-th layer.
While the discriminator $D$ is trained based on the following loss
\begin{equation}
\mathcal{L}^D_\mathrm{adv} = \mathbb{E}_{\mathbf{z}_\mathrm{in}, \mathbf{z}_\mathrm{tgt}} \left[ \left(1 - D(\mathbf{z}_\mathrm{tgt})\right)^2
+ 
D({F}_\theta(\mathbf{z}_\mathrm{in}))^2
\right].
\end{equation}

\subsection{Latent Space Variational Encoder for Mono-to-Stereo}
% --------------
Our second explored extension leverages the latent space to model a distribution over possible target outputs, rather than predicting a single deterministic result. While the reconstruction loss encourages predicting the mean of the target distribution, many audio tasks, such as spatial upmixing, have inherent ambiguity or require the ability to generate diverse, user-controllable outputs. To address this, we integrate a variational approach within the latent domain.

In this framework, our task-specific model ${F}_\theta$ is conditioned on a low-dimensional conditioning vector $\mathbf{c}\in\mathbb{R}^D$, produced by a separate encoder $G_\phi$ that maps the target latent $\mathbf{z}_{\mathrm{tgt}}$ to a conditional distribution over $\mathbf{c}$. This allows $\mathbf{c}$ to capture aspects of the target that are not strictly predictable from the input alone. Here $G_\phi$ predicts parameters $(\bm{\mu},\bm{\sigma})$ of a Gaussian and samples $\mathbf{c}$ via the reparameterization trick:
\begin{equation}
\mathbf{c} = \bm{\mu} + \bm{\sigma} \odot \bm{\epsilon}, \quad \bm{\mu},\bm{\sigma} = {G}_\phi(\mathbf{z}_\mathrm{tgt}), \quad \bm{\epsilon} \sim \mathcal{N}(\mathbf{0}, \mathbf{I}).
\end{equation}

We demonstrate this approach on mono-to-stereo (M2S) upmixing, which involves turning a mono waveform into a stereo one with realistic spatialization, creating a sense of width, depth, and directionality.
%This task is well-suited for a conditional generative approach as there are often multiple plausible stereo renderings for a given mono input, and controlling the spatial characteristics via a conditioning variable is desirable. 
To do this, ${F}_\theta$ takes the mono input latent ${\mathbf{z}}_\mathrm{in} \in \mathbb{R}^{C \times T}$ and $\mathbf{c}$ and is trained to predict the stereo left-right stacked target latent, $\hat{\mathbf{z}}_\mathrm{tgt} \in \mathbb{R}^{2 \times C \times T}$. The conditioning encoder, ${G}_\phi$, is fed the ground truth stereo target as input and performs a mean-pooling operation across time, outputting a single global $\mathbf{c}$. Both $F_\theta$ and $G_\phi$ are trained jointly with a $L_1$ reconstruction objective on the target stereo and a KL-regularization term on $(\bm{\mu},\bm{\sigma})$.

During inference, we perform blind upmixing by sampling the conditioning vector directly from the normal prior, $\mathbf{c} \sim \mathcal{N}(\mathbf{0}, \mathbf{I})$. By varying the sampled $\mathbf{c}$, we can generate diverse stereo outputs for the same mono input. This setup can, in principle, enable stereo style transfer by encoding a $\mathbf{c}$ from another stereo signal via $G_\phi$.
%This setup also enables stereo style transfer or specific spatial conditioning by encoding a $\mathbf{c}$ vector from a different stereo audio signal using the conditioning encoder $\mathrm{G}_\phi$ and then using that $\mathbf{c}$ to condition the prediction for the mono input. 

\section{Experimental Setup}
To evaluate our framework, we adapt a single recipe based on a shared pretrained and frozen autoencoder, and a common latent architecture.

\subsection{Models}

\subsubsection{Autoencoder} 
The base autoencoder is a variant of VOCOS~\cite{siuzdakvocos}, totaling 230~M parameters, using the same architecture for the encoder and decoder and downsampling by a factor of 1024. Its bottleneck is parameterized as a variational autoencoder with a latent representation of 64 channels and a 43~Hz latent frame rate. 

\subsubsection{Latent Modules} 
The latent modules employ a simple architecture consisting of sequential ConvNeXt-V2~\cite{woo2023convnext} blocks, preceded and followed by 1-by-1 Conv1d layers that appropriately adjust the number of channels. In every block, AdaLN introduces conditional information when needed. 
We construct two model variants of different sizes: the small (S) variant is comprised of $4$ blocks with a hidden dimension of $512$, resulting in 4.3~M parameters, while the medium (M) variant consists of $8$ blocks with a hidden dimension of $768$, totaling 19.1~M parameters. For BWE, we experiment with both variants. The mono-to-stereo model is based on the medium variant with a $2$-block conditioning branch, amounting to 24.8~M parameters. The stereo condition vector has $H=64$ dimensions.

The latent discriminator is designed to mimic a single-band of the multi-band discriminator~\cite{kumar2024high}. It consists of a 6-layer Conv2d stack with the first five using kernels of size $(3,7)$ and the last using $(3,3)$. All layers use hops and padding of $(1,1)$ and leaky relu nonlinearity. There is 1 input channel and the latent channels are all set to $256$.

\subsection{Training Details}

Models are trained with purely latent space losses as described in Sec.~\ref{sec:method} unless otherwise noted and in \texttt{bfloat16} precision. For BWE where we use the discriminator, we set the following loss weights: $10.0$ for the $\mathcal{L}_\mathrm{rec}$ term, $0.5$ for $\mathcal{L}^{F}_\mathrm{adv}$, and $1.0$ for $\mathcal{L}^{F}_\mathrm{fm}$. While, when training the mono-to-stereo model we use: $10.0$ for $\mathcal{L}_\mathrm{rec}$ loss and $5e-4$ for the KL loss term. Models are trained with a batch size of $256$ on audio chunks with a duration of 1.4~s for BWE, and 4~s for mono-to-stereo. We perform 250K training steps on a single H100-80GB GPU. We use AdamW with a learning rate of $5e-4$ for the main model and $1e-4$ for the latent discriminator, we also use linear warm-up totaling 1K steps for the main model and 20K steps for the latent discriminator. Weight decay is set at $1e-4$. 

\subsubsection{Data} We train on an internal dataset consisting of 10K~hours of genre and instrument diverse licensed instrumental music sampled at $44.1$~KHz. 
%We perform pitch-shifting and time-stretching augmentations using pyrubberband. 
In the case of BWE, we upsample $22.05$~KHz audio into $44.1$~KHz, as in~\cite{zhu2024musichifi}, we create paired data for training and evaluation by downsampling to $22.05$~KHz. BWE operates on mono audio. 

\subsection{Evaluation}

\subsubsection{Baselines}
For each task, we select specialized state-of-the-art (SOTA) baselines for comparison: MusicHiFi-BWE~\cite{zhu2024musichifi} and Aero~\cite{mandel2023aero} which outperform models like AudioSR~\cite{liu2024audiosr, zhu2024musichifi} for bandwidth extension, and MusicHiFi-M2S~\cite{zhu2024musichifi} for mono-to-stereo comparisons. In every task, we evaluate two scenarios: one where the baseline model has access to the undistorted input, and another involving autoencoder transmission where we first pass the model input through the autoencoder (denoted as $\mathrm{VAE + Baseline}$).

\subsubsection{Evaluation Data and Metrics}
We evaluate on the subset of FMA-small~\cite{defferrard2017fma} used in MusicHiFi~\cite{zhu2024musichifi}. As objective evaluation metrics, we use STFT and mel distance implemented by the auraloss library~\cite{steinmetz2020auraloss}. Reported FLOPS are measured by the calflops library. %Figures reported are based on inference performed with \texttt{float32} precision.

\section{Results}

\begin{figure}[t]
    \centering
     \vspace{0pt}
        \centering
       \includegraphics[width=0.95 \columnwidth,trim=10pt 15pt 10pt 10pt]{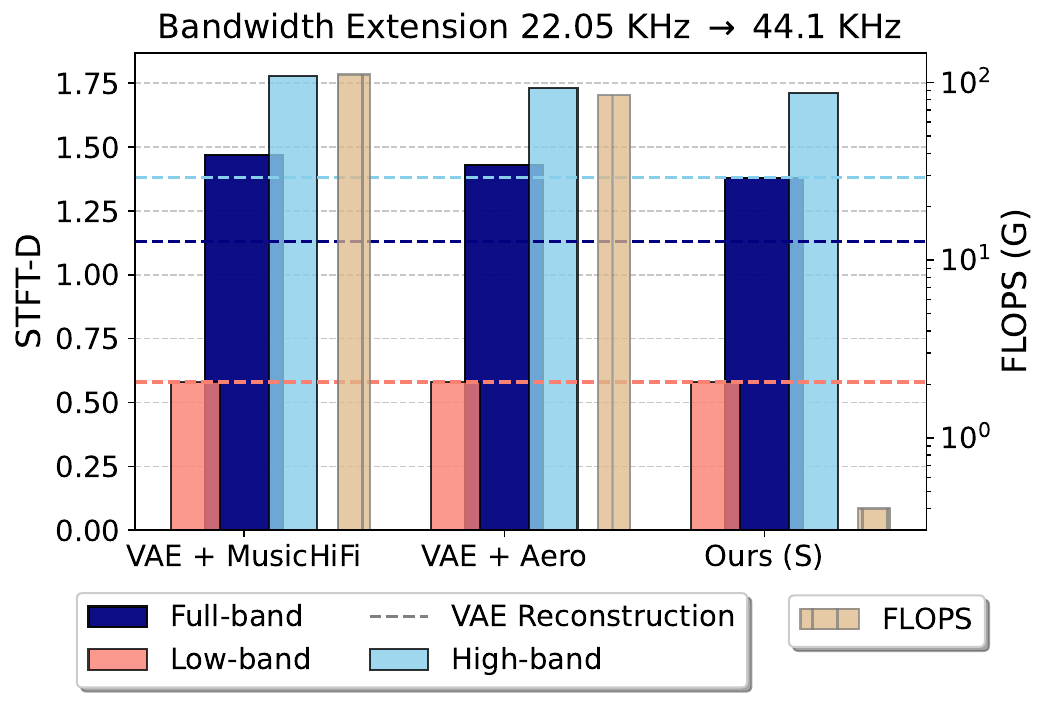}
        \caption{Left axis: Comparison of bandwidth extension methods in terms of STFT-distance~($\downarrow$) to the full-band targets. We also show low/high-band figures by filtering the two signals being compared with a low/high-pass filter respectively. Right axis: FLOPS required for processing one second of audio.}
    \label{fig:bwe}
\end{figure}

\begin{figure*}[t]
    \centering
    \includegraphics[width=\linewidth]{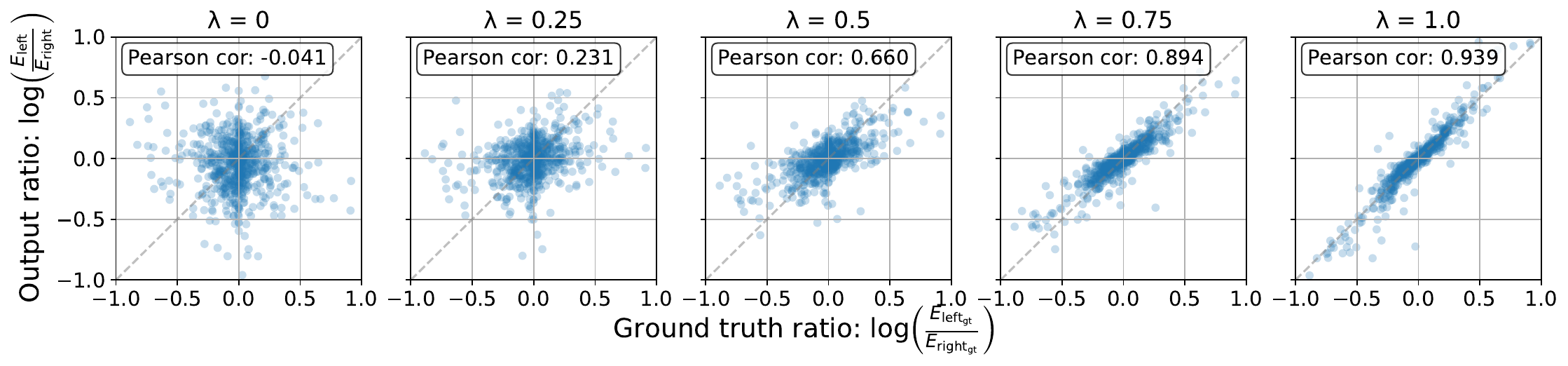}
    \caption{M2S control vector latent interpolation. The scatter plots show the relationship between the log-ratio of energy between the two channels in the ground truth stereo waveform $\log\left({E_{\text{left}_\mathrm{gt}}}/{E_{\text{right}_\mathrm{gt}}}\right)$ and the respective log-ratio $\log\left({E_{\text{left}}}/{E_{\text{right}}}\right)$ of the output stereo waveform for different values of $\lambda$.}
    \label{fig:m2s_latent_interpolation}
\end{figure*}

\subsection{Bandwidth Extension}

\begin{table}[t]
\caption{Bandwidth extension objective evaluation metrics on FMA-small. The row titled ``VAE rec." indicates the reconstruction error of the autoencoder on the full-band audios. ``VAE + Baseline" corresponds to baselines on a transmission scenario. Low/high-band results shown in parenthesis.}
\label{tab:bwe_vae_b}
\begin{center}
\resizebox{\columnwidth}{!}{  % 
\begin{tabular}{lccc}
\toprule
Model & GFLOPS & {STFT-D $\downarrow$} & {mel-D $\downarrow$} \\
\midrule
1 VAE rec.   &  51  & 1.13 (0.58 / 1.38) & 0.57 (0.46 / 0.76) \\
\midrule
2 Aero               & 85  & 0.94 (0.07 / 1.70) & 0.24 (0.03 / 0.76) \\
3 MusicHiFi-BWE      & 111 & 0.99 (0.09 / 1.74) & 0.24 (0.02 / 0.76) \\
\midrule
4 VAE + Aero               & 85  & 1.43 (0.58 / 1.73) & 0.67 (0.46 / 0.94) \\
5 VAE + MusicHiFi-BWE      & 111 & 1.47 (0.58 / 1.78) & 0.68 (0.46 / 0.98) \\
\midrule
6 Ours (M) with $L_1$            & 1.6 & 1.41 (0.58 / 1.78) & 0.69 (0.48 / 1.03) \\
7 Ours (M) with $L_1$, mel       & 1.6 & 1.35 (0.55 / 1.73) & 0.60 (0.42 / 0.92) \\
8 Ours (M) with $L_1$, Discr.    & 1.6 & 1.38 (0.58 / 1.72) & 0.67 (0.48 / 0.98) \\
9 Ours (S) with $L_1$, Discr.    & 0.4 & 1.38 (0.58 / 1.71) & 0.67 (0.48 / 0.97) \\
%Re-Encoder-L + $L_1$, discr     & 4.5 & 1.38 (0.58 / 1.71) & 0.67 (0.48 / 0.97) \\
\bottomrule
\end{tabular}
}
\end{center}
%\vspace{-1.5em}
\end{table}

This section describes experiments benchmarking latent processing and demonstrating the performance benefits of a latent discriminator. We focus on bandwidth extension and compare our approach against SOTA baselines: MusicHiFi~\cite{zhu2024musichifi} and Aero~\cite{mandel2023aero}. Unlike baselines operating on raw/spectral audio, our method processes VAE latent representations of low-band inputs. All models were evaluated within a simulated VAE transmission scenario with latent-compressed audio. In this setup, our proposed latent module processes the transmitted low-band latent representation directly. We verified that $A_D$ accurately reconstructs low-band signals. For the raw audio baselines, we simulate transmission by autoencoding the full-band signal and then generating low-band inputs. This evaluation framework is favorable to the baselines, as it assumes they operate on in-domain signals. Performance is quantified using STFT/mel -D metrics, overall and in low/high bands. Computational cost is measured in FLOPS. The full-band reconstruction error of the VAE (1.13 STFT-D, row 1) Table~\ref{tab:bwe_vae_b} serves as an empirical upper bound on performance in this scenario.

Results in Table~\ref{tab:bwe_vae_b} rows 2 and 3 show the performance of the raw audio baselines (Aero: 0.94 STFT-D, MusicHiFi: 0.99 STFT-D) when operating without a transmission constraint. These scores represent their empirical upper bound for the transmission scenario. The combinations of VAE + Aero (row 4) and VAE + MusicHiFi (row 5) yield STFT-D scores of 1.43 and 1.47, respectively. These results serve as our primary anchor point, allowing us to test whether latent upsampling can match or exceed the performance of post-hoc raw audio upsampling. Our `M' latent module with $L_1$ loss (row 6, 1.41) already surpasses these baselines in this transmission scenario. To understand the potential best performance achievable with a latent module, we trained an `M' variant with latent $L_1$ loss and decoder-propagated mel loss (row 7, 1.35). Remarkably, we are able to recover most of this strong mel-loss performance with a latent $L_1$ loss and a latent discriminator (row 8, 1.38), outperforming it in the high band (row 8 vs 7, 1.72 vs 1.73). This final latent model result is particularly surprising as it still surpasses our raw audio baselines (1.43, 1.47) despite operating on the latent space for both training and inference.

Finally, we compare computational efficiency. Our `S' latent module (row 9, 0.4 GFLOPS) requires approximately 200 times fewer FLOPS than the raw audio baselines (85 GFLOPS for Aero, 111 GFLOPS for MusicHiFi), demonstrating significant efficiency while achieving comparable performance with our `M' model.

\subsection{Mono-to-Stereo}
Next, we focus on mono-to-stereo conversion. Here, we benchmark latent processing and demonstrate the controllability benefits of incorporating a latent variational controller. We compare our approach against MusicHiFi~\cite{zhu2024musichifi}. Our method operates on mono-latent representations in VAE latent space; the baseline processes raw waveform audio. In a simulated VAE transmission scenario, our method processes the transmitted mono-latent directly, whereas the baseline operates on the VAE-decoded mono signal. Performance is quantified using STFT/mel -D (left/right and middle/side channel configurations), and computational cost is assessed using FLOPS. The VAE's intrinsic reconstruction error (1.14 left/right STFT-D, row 1, Table~\ref{tab:m2s_vae_b}) serves as an empirical upper bound on performance.

Table~\ref{tab:m2s_vae_b} row 2 shows the raw audio baseline's performance (0.85 STFT-D) without a transmission constraint. Row 3 presents the baseline performance with a transmission constraint (1.34 STFT-D left/right), operating on the decoded mono VAE output. MusicHiFi copies the mono input to the center and estimates a side channel. In contrast, our method estimates the left/right VAE latents directly, since side can be slightly out-of-domain for the VAE. We first evaluate our module by sampling the control vector from the prior, forgoing control. As shown in row 4, this latent module (1.31 STFT-D left/right) surpasses the transmitted raw audio baseline (1.34 STFT-D left/right) in left/right error. To demonstrate the potential of control, we provide the latent module with the oracle control vector, derived from the ground-truth stereo. In row 5 (1.22 STFT-D left/right), the controlled latent module improves performance for all metrics and formats with the exception of mel-D on side. Comparing computational costs, our latent module requires $138 \times$ fewer FLOPS than MusicHiFi. %We run a qualitative study on stereo control next.

We design a latent interpolation experiment to investigate the degree of spatial control provided by the conditional input $\mathbf{c}$. We interpolate between $\mathbf{c}_\mathrm{gt}$ the stereo representation of the ground truth stereo signal, and $\mathbf{c}_0$ sampled from the prior. Thus, the control vector is $\mathbf{c} = \lambda \mathbf{c}_\mathrm{gt} + (1 - \lambda) \mathbf{c}_0$, and our goal is to study the relationship between $\lambda$ and output spatial characteristics. Varying $\lambda \in [0,1]$ should transition between the two controls, with stereo reconstruction when $\lambda = 1$, a new stereo realization when $\lambda = 0$, and outputs with interpolated stereo characteristics in between. Results are shown in Fig.~\ref{fig:m2s_latent_interpolation} where each point corresponds to an audio sample, showing the channel energy ratios of the ground truth stereo and the output stereo realization given $\mathbf{c}$. As expected, for a well-behaved $\mathbf{c}$, the parameter $\lambda$ strongly influences the correlation between the two ratios. A value of $\lambda=0$ results in uncorrelated ratios, while $\lambda=1$ yields highly correlated ratios, indicating that the stereo reconstruction closely matches the ground truth stereo channel energy distribution. Interestingly, $\lambda$  exhibits a nearly linear relationship with the output correlation. This suggests that the conditional branch captures spatial cues, granting users control over output spatial characteristics.

% ------------------------------------

\begin{table}[t]
\caption{Mono-to-stereo objective evaluation metrics on FMA-small. ``VAE + MusicHiFi-M2S" corresponds to {MusicHiFi-M2S} on a transmission scenario. %with mono input autoencoded by VAE.
}
\label{tab:m2s_vae_b}
\begin{center}
\resizebox{\columnwidth}{!}{  % 
\begin{tabular}{lccccc}
\toprule
 & & \multicolumn{2}{c}{Left / Right} & \multicolumn{2}{c}{Middle / Side}  \\
Model                                   & GFLOPS & STFT-D $\downarrow$ & mel-D $\downarrow$ & STFT-D $\downarrow$  & mel-D $\downarrow$  \\
\midrule
1 VAE rec.                      & 51 & 1.14 / 1.14 & 0.58 / 0.58 & 1.13 / 1.48 & 0.57 / 0.85 \\
\midrule
2 MH-M2S                           & 222 & 0.85 / 0.85 & 0.62 / 0.62 & 0.00 / 2.18 & 0.00 / 1.89 \\
3 VAE + MH-M2S                     & 222 & 1.34 / 1.34 & 0.86 / 0.86 & 1.13 / 2.36 & 0.57 / 1.92 \\
\midrule
4 Ours (M) + Rand $\mathbf{c}$                               & 1.6 & 1.31 / 1.31 & 0.83 / 0.82 & 1.17 / 2.31 & 0.62 / 2.07 \\
5 Ours (M) + Oracle $\mathbf{c}$     & 1.6 & 1.22 / 1.22 & 0.71 / 0.72 & 1.13 / 2.24 & 0.58 / 2.19 \\
\bottomrule
\end{tabular}
}
\end{center}
%\vspace{-1.5em}
\end{table}

\section{Conclusion}
In this work, we introduced the Re-Encoder framework, an approach for performing audio processing operations entirely within the latent space of pre-trained neural audio autoencoders. This directly addresses the inherent inefficiency of operating on high-dimensional raw audio representations for audio-to-audio tasks. By leveraging these compact latents, our method dramatically simplifies the training process. It requires only a straightforward latent $L_1$ reconstruction loss, optionally augmented by a single latent adversarial discriminator or latent condition encoder. This stands in stark contrast to the complex multi-scale loss and multiple discriminator setups typically needed for conventional raw-audio methods. Through experiments in bandwidth extension and mono-to-stereo up-mixing, we demonstrated significant efficiency gains, achieving up to a $200\times$ speedup while maintaining output quality. By keeping operations entirely within the efficient latent domain, the Re-Encoder establishes a more efficient paradigm for audio processing pipelines which use autoencoders during both training and inference. This paves the way for faster, more resource-efficient workflows across a variety of audio tasks.

% \section{Acknowledgment}

% We would like to thank Nick Bryan, Juan-Pablo Caceres, and Zhepei Wang for their insightful discussions and feedback. We would also like to thank Ge Zhu for providing the Music-HiFi checkpoints.

\clearpage
% The \IEEEtriggeratref{XX} command can be used to move to the next column before the XX-th reference
% to balance the two columns of the reference section
% \IEEEtriggeratref{XX}
\bibliographystyle{IEEEtran}
\bibliography{refs25}
% or list them by yourself:
% \begin{thebibliography}{1}

\end{document}